\documentclass[aps,showpacs,amsmath,amssymbd,dvips,showkeys,preprintnumbers]{revtex4}

\usepackage{longtable}
\usepackage{supertabular}
\usepackage{float}
\usepackage{tikz}
\usetikzlibrary{arrows,shapes,positioning}
\usepackage{pgfplots}
\usetikzlibrary{arrows}
\usepackage{relsize}
\usepackage{lscape}
\usepackage{graphicx}
\usepackage{color}

\def \be  {\begin{equation}}
\def \ee  {\end{equation}}
\def \ee  {\end{equation}}
\def \bea {\begin{eqnarray}}
\def \eea {\end{eqnarray}}

\newcommand{\nn}{\nonumber}

\begin{document}

\preprint{ECTP-2016-10}
\preprint{WLCAPP-2016-10}
\hspace{0.05cm}
\title{SU($3$) Polyakov linear-sigma model: bulk and shear viscosity of QCD matter in finite magnetic field}

\author{Abdel Nasser  TAWFIK}
\email{a.tawfik@eng.mti.edu.eg}
\affiliation{Egyptian Center for Theoretical Physics (ECTP), Modern University for Technology and Information (MTI), 11571 Cairo, Egypt}
\affiliation{World Laboratory for Cosmology And Particle Physics (WLCAPP), 11571 Cairo, Egypt}

\author{Abdel Magied DIAB}
\affiliation{Egyptian Center for Theoretical Physics (ECTP), Modern University for Technology and Information (MTI), 11571 Cairo, Egypt}
\affiliation{World Laboratory for Cosmology And Particle Physics (WLCAPP), 11571 Cairo, Egypt}

\author{M. T. Hussein}
\affiliation{Physics Department, Faculty of Science,  Cairo University, 12613 Giza, Egypt}

\begin{abstract}
Due to off-center relativistic motion of the charged spectators and the local momentum-imbalance of the participants, a short-lived huge magnetic field is likely generated, especially in relativistic heavy-ion collisions. In determining the temperature  dependence of bulk and shear viscosities of the QCD matter in vanishing and finite magnetic field, we utilize mean field approximation to the SU($3$) Polyakov linear-sigma model (PLSM). We compare between the results from two different approaches; Green-Kubo correlation and  Boltzmann master equation with Chapman-Enskog expansion. We find that both approaches have almost identical results, especially in the hadron phase. In the temperature dependence of bulk and shear viscosities relative to thermal entropy at the critical temperature, there is a rapid decrease in the chiral phase-transition and in the critical temperature with increasing magnetic field. As the magnetic field strength increases, a peak appears at the critical temperature ($T_c$). This can be understood from the  small drop on the thermal entropy at $T_c$, which can be interpreted due to instability in the hydrodynamic flow of the quark-gluon plasma and soft statistical hadronization. It is obvious that, increasing magnetic field accelerates the transition from hadron to QGP phases (inverse catalysis), i.e., taking place at lower temperatures.

\end{abstract}


\maketitle

\tableofcontents
\makeatletter
\let\toc@pre\relax
\let\toc@post\relax
\makeatother

\section{Introduction \label{intro}}

Recently, the study of the influence of strong magnetic field on Quantum Chromodynamics (QCD) apparently gains increasing popularity among particle physicists.  Such a strong magnetic field can be reproduced in various high-energy regimes such as early universe and non-central heavy-ion collisions (HIC) \cite{Skokov:2009,Vachaspati:1991}. In the heavy-ion experiments, a huge magnetic field can be created due to the relativistic motion of charged spectators and the local momentum-imbalance of the participants. At SPS, RHIC and LHC energies, the expected magnetic field ranges between $0.1\, m_{\pi}^2$, $m_{\pi}^2 $ and $10-15 \, m_{\pi}^2$, respectively \cite{Skokov:2009, Elec:Magnet}, where $m_{\pi}^2 \sim 10^8$ Gauss. 

The influence on QCD doesn't only cause catalysis of the chiral symmetry breaking \cite{Bruckmann:2013, Preis:2011} but also modifies the chiral phase structure of the hadron production. Also, it changes the nature of the chiral phase-transition \cite{Sanfilippo:2010, catalysis:2014, Catalysis:2015} and the energy loss due to quark synchrotron radiation \cite{Synchrotron:2010, Elec:Magnet}. Furthermore, the magnetic field does not only come up with essential effects during the early stages of HIC, but also during the later ones, where the response of the magnetic effect is assumed to have a large in-medium-dependence. The latter depends on the variation of the magnetic diffusion time \cite{Synchrotron:2010, Elec:Magnet} and the electrical conductivity which are medium depending \cite{Gupta:2004,Bratkovskaya:1995}. 

The description of the chiral and deconfinement phase-structure of the hadrons, the characterization of the QGP properties and the definition of the critical endpoint (CEP)  are examples on significant researches conducted during last decades. The transport properties are particularly helpful in characterizing strongly interacting QCD matter, such as the phase transition, the critical endpoint, etc. \cite{Kapusta:1993}. The viscous transport properties have been reviewed in Ref. \cite{Kapusta:2008}. The response of the QCD matter to an external magnetic field can be described by the transport coefficients, such as bulk and shear viscosities. In the present study, we extend our previous work \cite{TD:trans}, where the temperature dependence of bulk and shear viscosities was deduced from SU($3$) PLSM to a finite magnetic field \cite{Kubo:1957}. The bulk [$\zeta(T,eB)$]  and shear [$\eta(T,eB)$] viscosity normalized to the entropy density $s(T,eB)$ shall be calculated at finite temperatures and magnetic field strengths. We also address the chiral and deconfinement phase-transitions in finite magnetic field.

First, we recall that so-far various LSM-calculations have been performed in order to determine the viscous properties of the QCD matter  \cite{Dobado:2009,Dobado:2012,Chakraborty:2011}. Based on Boltzmann-Uehling-Uhlenbeck (BBU) equation and Green-Kubo (GK) correlation, $\eta/s$ has been estimated in the large-N limit \cite{Dobado:2009}. Also, $\zeta/s$ in the large-N limit has been calculated from Boltzmann-Uehling-Uhlenbeck \cite{Dobado:2012}. From relaxation time approximation (RTA) and BUU equation, the shear and bulk viscosity have been calculated in SU($2$) LSM \cite{Chakraborty:2011}. Second, from BUU equation with relaxation time approximation, some of such dissipative properties haven been studied from the hadron resonance gas (HRG) model with excluded-volume corrections as function of temperature and baryon chemical potential \cite{Kadam:2015}.

In the present work, it is assumed that the temperature dependence of  QCD viscous properties such as bulk and shear viscosity are strongly affected by the huge short-lived magnetic field, which can be generated in relativistic heavy-ion collisions. We study their dependence on various magnetic field strengths. We present a direct estimation for both types of viscosity coefficients from PLSM by using BUU and GK approaches. For the first time, a systematic study in SU($3$) PLSM in vanishing and nonzero magnetic field is presented. Such a way we can compare between the results from these two different approaches. A rapid decrease in the chiral phase-transition and in the critical temperature with increasing magnetic field is observed. Increasing magnetic field is accompanied by phase transitions that take place at lower critical temperatures relative to the ones at vanishing magnetic fields.  In other words, increasing magnetic field leads to a decrease in the corresponding critical temperature (inverse catalysis).

This paper is organized as follows, we briefly describe PLSM in mean field approximation in section \ref{model} in which information about hadron matter in the presence of magnetic field is included. BUU and GK approaches are introduced in section \ref{2approaches} and elaborated in Appendices \ref{VBUU6} and \ref{VGK0}, respectively. The temperature dependence of the relaxation time and the bulk and shear viscosities normalized to the thermal entropy at finite magnetic field strength and vanishing chemical potential shall be elaborated in section \ref{viscosity}. This is followed by the conclusions in section \ref{conclusion}.

\section{Reminder to SU($3$) linear-sigma model with mean field approximation \label{model}}

The exchange of energy between particle and antiparticle at temperature ($T$) and baryon chemical potential ($\mu_f$) can be included in the grand canonical partition function ($\mathcal{Z}$),
\begin{eqnarray}
\mathcal{Z}
&=& \int\prod_a \mathcal{D} \sigma_a \mathcal{D} \pi_a \int
\mathcal{D}\psi \mathcal{D} \bar{\psi} \mathrm{exp} \left[ \int_x
(\mathcal{L}+\sum_{f} \mu_f \bar{\psi}_f \gamma^0 \psi_f )
\right],
\end{eqnarray} 
where $\int_x\equiv i \int^{1/T}_0 dt \int_V d^3x$ and $V$ is the volume of the system of interest. The subscript $f$ refers to quark flavors and therefore $\mu_f$ is the chemical potential for quark flavors $f=(l,s,\bar{l},\bar{s})$. One can define a uniform blind chemical potential $\mu_f \equiv \mu_{u, d} = \mu_s$ \cite{blind,Schaefer:2007c,Schaefer:2008hk} as a result of the assumption of symmetric quark matter and degenerate light quarks. $\mathcal{L}$ is a Lagrangian coupled the chiral LSM Lagrangian with the Polyakov loops potential, $\mathcal{L}=\mathcal{L}_{\mbox{chiral}}-\mathbf{\mathcal{U}} \left(\phi, \phi^*, T\right)$. More details about the PLSM model can be found in Refs. \cite{TD:Masses,TND:2014,TN:magnet}. Moreover, the free energy can be given as   $\mathcal{F}=-T \cdot \log [\,\mathcal{Z}]/V$  or 
\begin{eqnarray}
 \mathcal{F}  
&=&  U(\sigma_l, \sigma_s) +\mathbf{\mathcal{U}}(\phi, \phi^*, T) + \Omega_{\bar{q}q}(T, \mu _f, B)  + \delta_{0,eB} \,\Omega_{ \bar{q}q}(T, \mu _f). \label{potential}
\end{eqnarray}

\begin{itemize}
\item The purely mesonic potential is given as
 \begin{eqnarray}
U(\sigma_l, \sigma_s) &=& - h_l \sigma_l - h_s \sigma_s + \frac{m^2}{2}\, (\sigma^2_l+\sigma^2_s) - \frac{c}{2\sqrt{2}} \sigma^2_l \sigma_s  
+ \frac{\lambda_1}{2} \, \sigma^2_l \sigma^2_s +\frac{(2 \lambda_1 +\lambda_2)}{8} \sigma^4_l  + \frac{(\lambda_1+\lambda_2)}{4}\sigma^4_s . \hspace*{8mm} \label{Upotio}
\label{pure:meson}
\end{eqnarray}

\item  In the present work, we implement the polynomial form of the Polyakov loop potential \cite{Ratti:2005jh, Schaefer:2007d, Roessner:2007, Fukushima:2008wg},  
\begin{eqnarray}
\frac{\mathbf{\mathcal{U}}\left(\phi, \phi^*, T\right)}{T^4}=-\frac{b_2(T)}{2}\left(\left|\phi\right|^2 + \left|\phi ^*\right|^2\right)-\frac{b_3
}{6}\left(\phi^3+\phi^{*3}\right)+\frac{b_4}{16}\left(\left|\phi\right|^2 +\left|\phi^*\right|^2\right)^2, \label{Uloop}
\end{eqnarray}
where $b_2(T)=a_0+a_1\left(T_0/T\right)+a_2\left(T_0/T\right)^2+a_3\left(T_0/T\right)^3$. With the parameters $a_0=6. 75$, $a_1=-1. 95$,  $a_2=2. 625$,  $a_3=-7. 44$, $b_3 = 0.75$ and $b_4=7.5$ \cite{Ratti:2005jh}, the pure gauge QCD thermodynamics is well reproduced. For a better agreement with lattice QCD simulations, the critical temperature $T_0$ is fixed at $187~$MeV for $N_f=2+1$ \cite{Schaefer:2007d}. 

\item The quarks and antiquark contribution to the medium potential can be divided into two regimes.
\begin{itemize}
\item In vanishing magnetic field ($eB=0$) but at finite $T$ and $\mu_f$ \cite{kapusta1989}, 
\begin{eqnarray} 
\Omega_{  \bar{q}q}(T, \mu _f)&=& -2 \,T \sum_{f} \int_0^{\infty} \frac{d^3\vec{p}}{(2 \pi)^3} \; f_f (T,\mu).
\end{eqnarray}
When introducing Polyakov-loop corrections to the quark's degrees of freedom, then the quark Fermi-Dirac distribution function becomes 
\begin{eqnarray}
f_f (T,\mu) &=&\ln \left[ 1+3\left(\phi+\phi^* \,e^{-\frac{E_f-\mu _f}{T}}\right)\times e^{-\frac{E_f-\mu _f}{T}}+e^{-3 \frac{E_f-\mu _f}{T}}\right], \label{fqaurk} 
\end{eqnarray}
where $E_f =(m_f^2 +p^2)^{1/2}$ is the dispersion relation of $f$-th quark flavor. For antiquarks, $\phi$ and $\phi^*$ are replaced with each other and the chemical potential $-\mu$ should be replaced by $\mu$.

\item In nonzero magnetic field  ($eB\neq 0$) but at finite $T$ and $\mu_f$,  the concepts of Landau quantization and magnetic catalysis, where the magnetic field is assumed to be oriented along $z$-direction, should be implemented.  According to the magnetic catalysis \cite{Shovkovy:2013},
\bea
 \int \frac{d^3p}{(2\pi)^3}  \longrightarrow   \frac{|q_f|B}{2\pi} \sum_\nu \int \frac{dp_z}{2\pi} (2-\delta_{0\nu}), \label{phaseeB}
\eea
\begin{eqnarray} 
\Omega_{  \bar{q}q}(T, \mu _f, B)&=& - 2 \sum_{f} \frac{|q_f| B \, T}{(2 \pi)^2} \,  \sum_{\nu = 0}^{\infty}  (2-\delta _{0 \nu })    \int_0^{\infty} dp_z \; f_{f} (T, \mu,eB). \label{PloykovPLSM}
\end{eqnarray}
The distribution function in finite magnetic field can be given as
\bea
f_{f} (T, \mu,eB) &=&\ln \left[ 1+3\left(\phi+\phi^* e^{-\frac{E_{B, f} -\mu _f}{T}}\right)\; e^{-\frac{E_{B, f} -\mu _f}{T}} +e^{-3 \frac{E_{B, f} -\mu _f}{T}}\right].
\eea
For antiquarks, a similar expression can derived. It is noteworthy highlighting that the dispersion relation in nonzero magnetic field gets modification as follows. 
\bea
E_{B, f}&=&\left[p_{z}^{2}+m_{f}^{2}+|q_{f}|(2n+1-\sigma) B\right]^{1/2}. \label{eq:moddisp}
\eea 
The quantization number ($n$) is known as the Landau quantum number $\nu$. $\sigma$ is related to the spin quantum number, $\sigma=\pm S/2$ and to the masses of quark-flavor $f=l,s$ with $l$ runs over $u$ and $d$ quarks and the other subscript stands for $s$-quarks. For the latter, the massed are directly coupled to the sigma fields
\bea
m_l = g\, \frac{\sigma_l}{2}, \qquad & & \qquad
m_s = g\, \frac{\sigma_s}{\sqrt{2}}.  \label{qmassSigma}
\eea
We note that the quantity $2n+1-\sigma$ can be replaced by sum over the Landau Levels. For completeness, we mention that $2-\delta_{0 \nu}$ represents degenerate Landau Levels. 
\end{itemize}
\end{itemize}

When assuming global minimization of the free energy ($\mathcal{F}$), 
\begin{eqnarray}
\left.\frac{\partial \mathcal{F} }{\partial \sigma_l} = \frac{\partial
\mathcal{F}}{\partial \sigma_s}= \frac{\partial \mathcal{F} }{\partial
\phi}= \frac{\partial \mathcal{F} }{\partial \phi^*}\right|_{min} &=& 0, \label{cond1}
\end{eqnarray}
the remaining parameters $\sigma_l=\bar{\sigma_l}$, $\sigma_s=\bar{\sigma_s}$, $\phi=\bar{\phi}$ and $\phi^*=\bar{\phi^*}$ and their dependences on $T$, $\mu$ and $e B$ can be determined.

\section{Approaches \label{2approaches}}

\subsection{Boltzmann-Uehling-Uhlenbeck (BUU) equation}
\label{sec:buu}

From relativistic kinetic theory, the transport coefficients of the system of interest can be estimated in non-Abelian external field. At finite baryon (fermion) density, the relaxation time approximation can be applied to the Boltzmann-Uehling-Uhlenbeck (BUU) equation  \cite{Chakraborty:2011} with Chapman-Enskog expansion. The Bulk and shear viscosities are given as \cite{Chakraborty:2011},
\bea
\zeta (T,\mu) &=&  \frac{1}{9T}  \sum_{f} \int \frac{d^3p}{(2\pi)^3} \, \frac{\tau _f}{E_f ^2}\, \left[\frac{|\vec{p}|^2}{3} - c_s^2 E_f^2 \right]^2 \, f_f (T, \mu), \\ \nonumber
\eta (T,\mu) &=&  \frac{1}{15T}  \sum_{f} \int \frac{d^3p}{(2\pi)^3}\, \frac{p^4}{E_f^2}\,  \tau _f  f_f (T, \mu).
\eea

In a nonzero magnetic field ($eB\ne 0$), it is convenient to derive the relaxation time approximation formulas
for bulk and shear viscosity. We start with BUU  and Chapman-Enskog expansion. More details are elaborated in Appendix \ref{VBUU6}. The bulk and shear viscosities read
\bea
\zeta (T, \mu, eB) &=&  \frac{1}{9\, T}  \sum_f  \frac{|q_f|B}{2\pi} \sum_\nu \int \frac{dp}{2\pi} \left(2-\delta_{0\nu}\right) \, \frac{\tau_f}{E_{B,f} ^2} \, \left[\frac{|\vec{p}|^2}{3} - c_s^2 E_{B,f}^2 \right]^2 \, f_{f} (T, \mu) ,  \label{eq:buubulk} \\
\eta (T,\mu, eB) &=&  \frac{1}{15\, T}  \sum_f  \frac{|q_f|B}{2\pi} \sum_\nu \int \frac{dp}{2\pi} \left(2-\delta_{0\nu}\right)\, \frac{p^4}{E_{B,f}^2}\,  \tau_f  f_f (T, \mu).  \label{eq:buushear}
\eea

\subsection{Green-Kubo (GK) correlation \label{sec:gk}}

Corresponding to dissipative fluxes, the Green-Kubo (GK) correlation, which is based on the linear response theory (LRT)  \cite{Fraile:2009,Fraile:2007}, directly relates the transport coefficients to out and in equilibrium correlation. The dissipative fluxes are treated as perturbations to the local thermal equilibrium. In doing this, the transport coefficients associated with the conserved quantities can be formulated as the expected values at equilibrium \cite{Fraile:2009,Fraile:2007}. The lowest order contribution to bulk and shear viscosity, respectively \cite{Fraile:2009,Fraile:2007} are given as
\bea
\zeta (T, \mu) &=& \frac{3}{2T} \sum_f  \int \, \frac{d^3p}{(2\pi)^3} \, \frac{\tau_f}{E_f^2} \left[\frac{|\vec{p}|^2}{3} - c_s^2\, E_f^2 \right]^2 \, f_{f} (T, \mu) \Big[1 + f_{f} (T, \mu)  \Big], \\
\eta (T, \mu) &=& \frac{2}{15T}  \sum_f \int \, \frac{d^3p}{(2\pi)^3} \, \frac{|\vec{p}|^4 \tau_f}{E_f^2} \, \, f_{f} (T, \mu) \Big[1 + f_{f} (T, \mu)  \Big],
\eea
where the Fermi-Dirac distribution function for $f$-th quark flavor $f_{f} (T, \mu)$ is given by Eq. (\ref{fqaurk}). 

In a nonzero magnetic field $eB\ne 0$ and by using LRT (diagrammatic approach), Appendix \ref{VGK0}, the bulk and shear viscosity, respectively, can be given as
\bea
\zeta (T,\mu, eB) &=&  \frac{3}{2\, T}   \sum_f  \frac{|q_f|B}{2\pi} \sum_\nu \int \frac{dp}{2\pi}  \left(2-\delta_{0\nu}\right) \,  \frac{\tau_f}{E_{B,f}^2} \left[\frac{|\vec{p}|^2}{3} - c_s^2 E_{B,f}^2 \right]^2 \, f_{f} (T, \mu,eB) \Big[1 + f_{f} (T, \mu,eB) \Big], \label{eq:gkbulk}\\
\eta (T,\mu, eB) &=& \frac{2}{15\, T} \,  \sum_f  \frac{|q_f|B}{2\pi} \sum_\nu \int \frac{dp}{2\pi}  \left(2-\delta_{0\nu}\right) \,  \frac{|\vec{p}|^4 \tau_f}{E_{B,f}^2} \; f_{f} (T, \mu,eB) \Big[1 + f_{f} (T, \mu,eB)  \Big]. \label{eq:gkshear}
\eea

\section{Results \label{viscosity}}

\subsection{Quark relaxation time}
\label{sec:rt}

In order to compute bulk and shear viscosities from BUU or GK approaches, Sec. \ref{sec:buu} and Sec. \ref{sec:gk}, respectively, a reliable estimation for the relaxation time ($\tau_f$) is very essential. In framework of PLSM, the quark flavors represent the effective degrees of freedom, especially at high temperatures. Thus, the relaxation time of such a quark system is what we need to estimate for the present work. At low temperatures, the hadronic degrees of freedom, pion and sigma mesons, become dominant.

\begin{figure}[htb]
\centering{
\includegraphics[width=5.5cm,angle=-90]{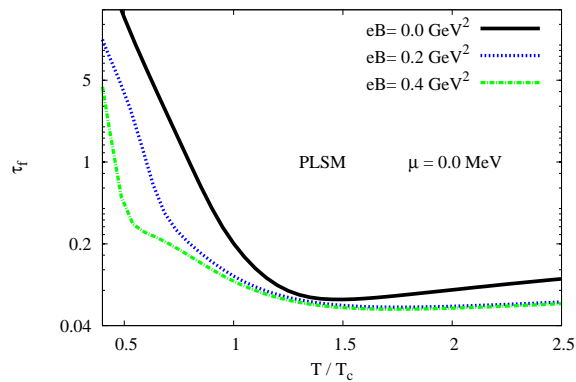}
\caption{\footnotesize (Color online) The relaxation time of $f$-th quark flavor ($\tau_f$) is calculated from PLSM in dependence on temperature at vanishing baryon chemical potential and different magnetic field strengths $eB=0.0~$GeV$^2$ (solid) $eB=0.2~$GeV$^2$ (dotted) and $eB=0.4~$GeV$^2$ (dot-dashed curve). \label{fig:taurelaxation}
}}
\end{figure}

For a microscopic consideration, the relaxation time can be determined from the thermal average of total elastic scattering and depends on the relative cross section $\sigma_{tr} (T) $,
\bea
\tau = \left[{n_f\, \langle \upsilon_{rel} (T) \, \sigma_{tr} (T) \rangle}\right]^{-1},
\eea 
where $\langle\upsilon_{rel}\rangle$ is the mean relative velocity between the two colliding particles and $n_f$ is their number density. 

In relativistic kinetic theory, the shear viscosity normalized to thermal entropy ($\eta/s$) likely remains unchanged due to the dynamics of the collisions \cite{Greco:2009}. In local spacetime coordinates, this ratio gives an estimation for  the strength of the cross section $\sigma_{tr}$ in $i$-th cell \cite{Groot:1980} 
\bea
\sigma_{tr,i}(T) &=& \frac{4}{15} \frac{\langle p\rangle_i }{\rho_i (4 - \mu_i/T)}\, \frac{1}{\eta/s}, \label{eq:sigmaT}
\eea
with $4\pi \eta/s$ sets in the range between $1$ and $4$ and $\rho_i$ is energy density. For the sake of simplicity, the temperature dependence of $\sigma_{tr,i}$ can be determined from a free massless gas, which is likely related to relativistic collisions. In this limit, the entropy is given by $s/T^3=g_f\, (2\, \pi^2)/45$. At vanishing $\mu_i$, then $\sigma_{tr} \sim T^{-2}$ \cite{Greco:2009}. Furthermore, from Bjorken picture \cite{Armesto:2008,Molnar:20089}, $T \sim \tau^{-1/3}$, $\sigma_{tr} \approx \tau^{2/3}$ and the cross section $\sigma_{tr} \sim T^{-2}$. In light of this, the relaxation time can  approximately be determined from PLSM number density. Its temperature evolution is thus very obvious. The density dependence requires to keep $\mu_i$ finite in Eq. (\ref{eq:sigmaT}). The present work, in contrary, assumes vanishing chemical potential.

In Fig. \ref{fig:taurelaxation}, a numerical estimation for the relaxation time of $f$-th quark flavor ($\tau_f$) in a wide range of temperature and magnetic field strengths $eB=0.0~$GeV$^2$ (solid), $eB=0.2~$GeV$^2$ (dotted) and $eB=0.4~$GeV$^2$ (dot-dashed curve) is depicted. It is obvious that increasing the magnetic field strength lowers the relaxation time, especially at low temperatures. In other words, the stronger becomes the magnetic field strength the slower is the temperature dependence of the relaxation time. In this temperature limit, $\tau_f$ almost exponentially decreases with the temperature. At high temperatures, the relaxation time becomes nearly temperature independent, regardless a very slow increase in $\tau_f$ is observed with increasing temperature.

\subsection{Bulk and shear viscosities from BUU and GK}
\label{sec:resVisc}

\begin{figure}[htb]
\centering{
\includegraphics[width=5.5cm,angle=-90]{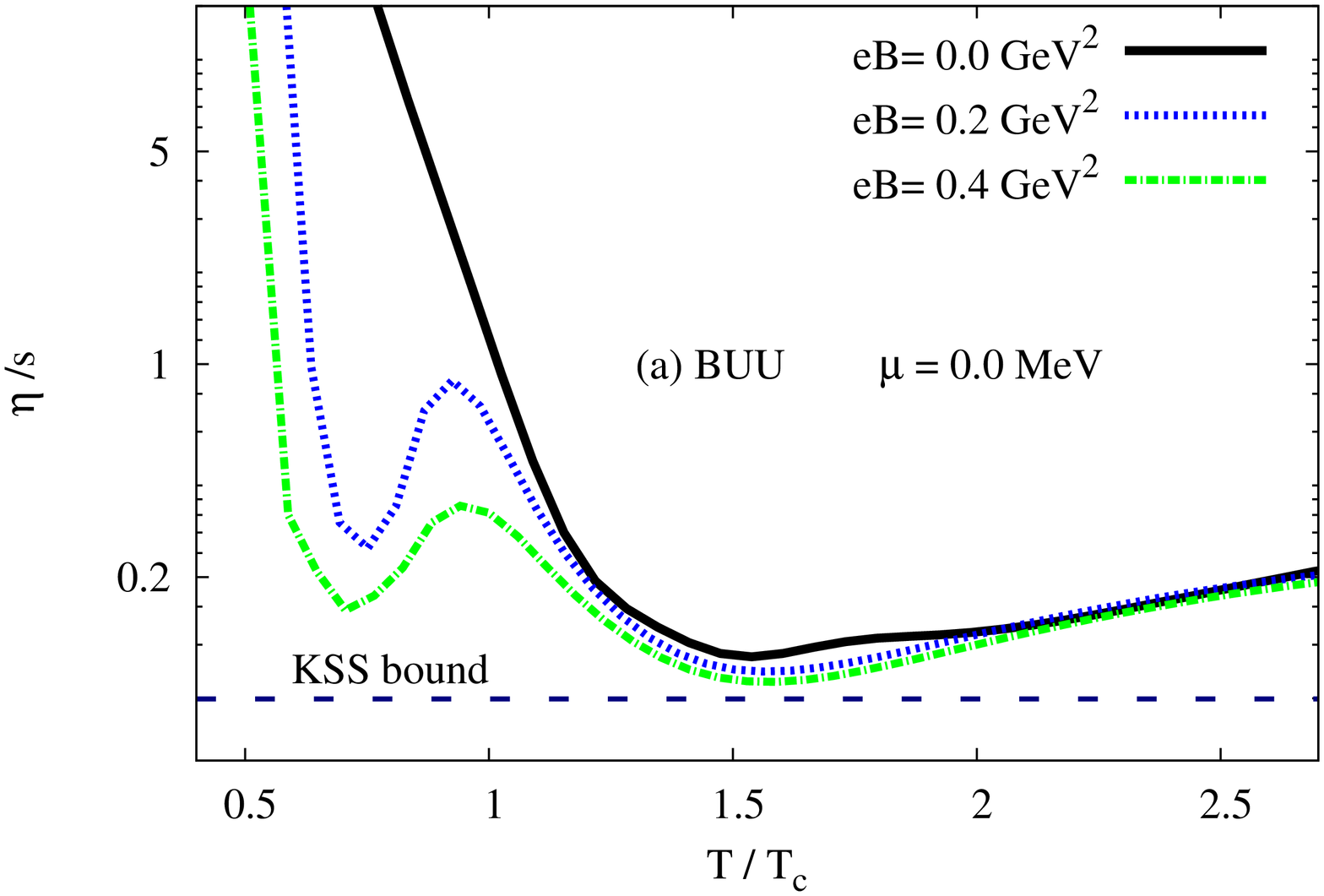}
\includegraphics[width=5.5cm,angle=-90]{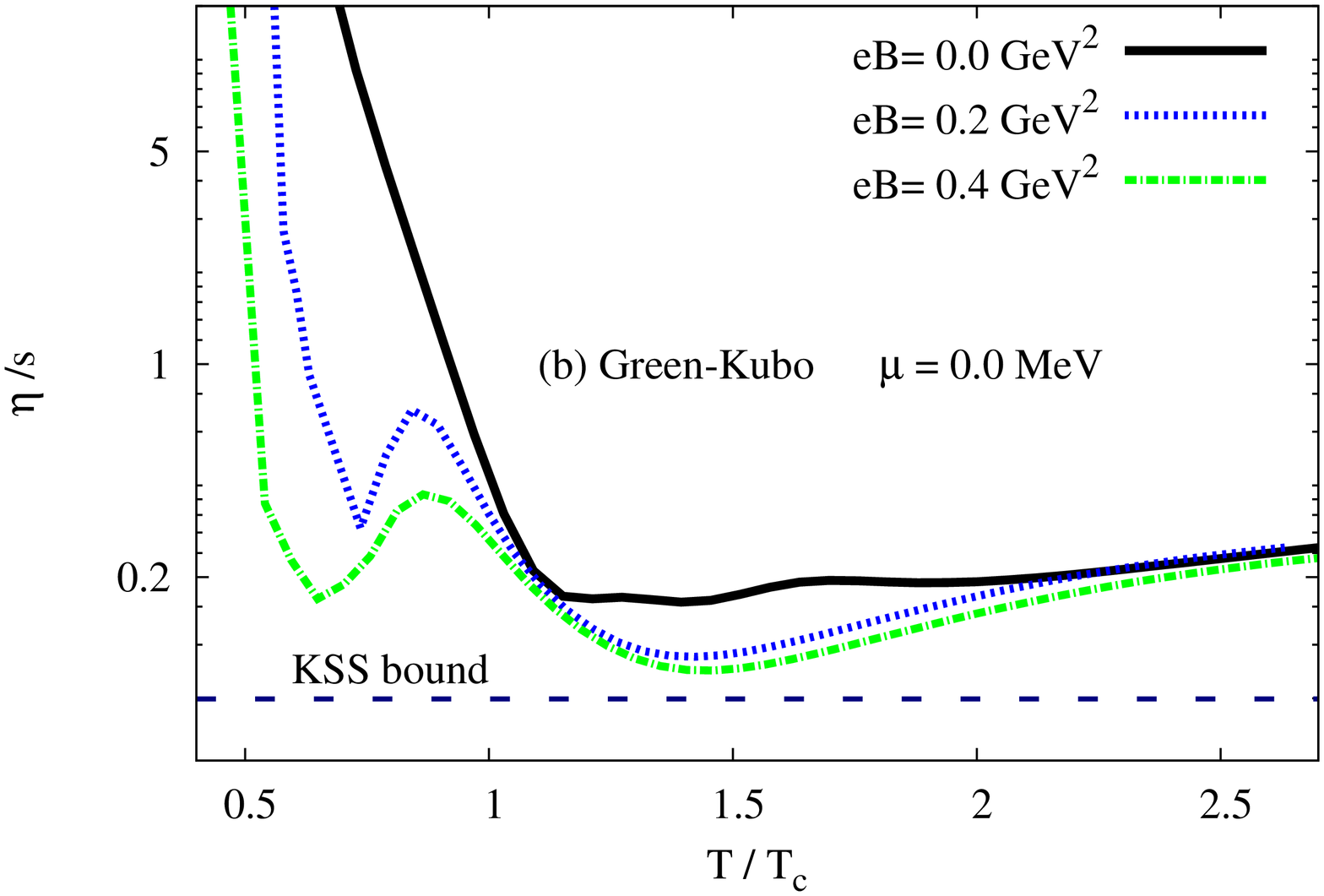}
\caption{\footnotesize (Color online) $\eta/s$ calculated from PLSM at different magnetic filed strengths $eB=0.0~$GeV$^2$ (solid), $eB=0.2~$GeV$^2$ (dotted) and $eB=0.4~$GeV$^2$ (dot-dashed) and at vanishing chemical potential is given as a function of temperature. Left-hand panel (a) shows the results form Boltzmann-Uehling-Uhlenbeck equation, while the right-hand panel (b) presents the results from Green-Kubo correlation. \label{fig:shear_viscosity2}
}}
\end{figure}

Fig. \ref{fig:shear_viscosity2} depicts the magnetic field effects on the temperature dependence of $\eta/s$ at vanishing chemical potential. The solid curve presents the results at a vanishing magnetic field, while the results at $eB=0.2$ and $0.4~$GeV$^2$  are given as dotted  and dot-dashed curves, respectively. The left-hand panel (a) shows $\eta/s$ as calculated from Boltzmann-Uehling-Uhlenbeck equation and the right-panel presents the calculations from GK correlation. The Kovtun, Son, and Starinets (KSS) limit is represented by dashed line. 

It is obvious that both approaches give almost identical $\eta/s$ values. Their temperature dependence is almost similar. Almost same results have been reported in Ref. \cite{Dobado:2009}. The ratio $\eta/s$ starts from a very large value at low temperature. Increasing temperature almost exponentially decreases $\eta/s$. But at high temperature, there is a small increase observed with increasing temperature. In nonzero magnetic field, there is an obvious enhancement in the rapid decrease relative to its values at low temperature. Furthermore, increasing the magnetic field strength makes the temperature-dependence more steeply. It is worthwhile to notice the appearance of  characterizing peaks at the critical temperature. Such peaks are connected with minima at lower temperatures. At high temperature, there is a slight increase in $\eta/s$ with increasing temperature. Furthermore, we notice that the resulting $\eta/s$ seems not depending on $eB$. Also, we notice that our numerical estimations for $\eta/s$ from PLSM is larger than KSS limit.

Some remarks on the peaks that characterize the phase transition are in order now. The magnetic field is believed to keep some effects from the hadronic phase  and affects the particle production and the deconfinement \cite{LatticeBali}. Accordingly, the peaks seems to favor two different scenarios. The first one is the instability in the hydrodynamic flow of QGP \cite{Torrieri:2009}. The second one is the soft statistical hadronization \cite{Brasoveanu:2010,Karsch:2008}. At high temperatures, the QCD coupling become weak and the hadrons are entirely liberated into quarks and gluons.

\begin{figure}[htb]
\centering{
\includegraphics[width=5.5cm,angle=-90]{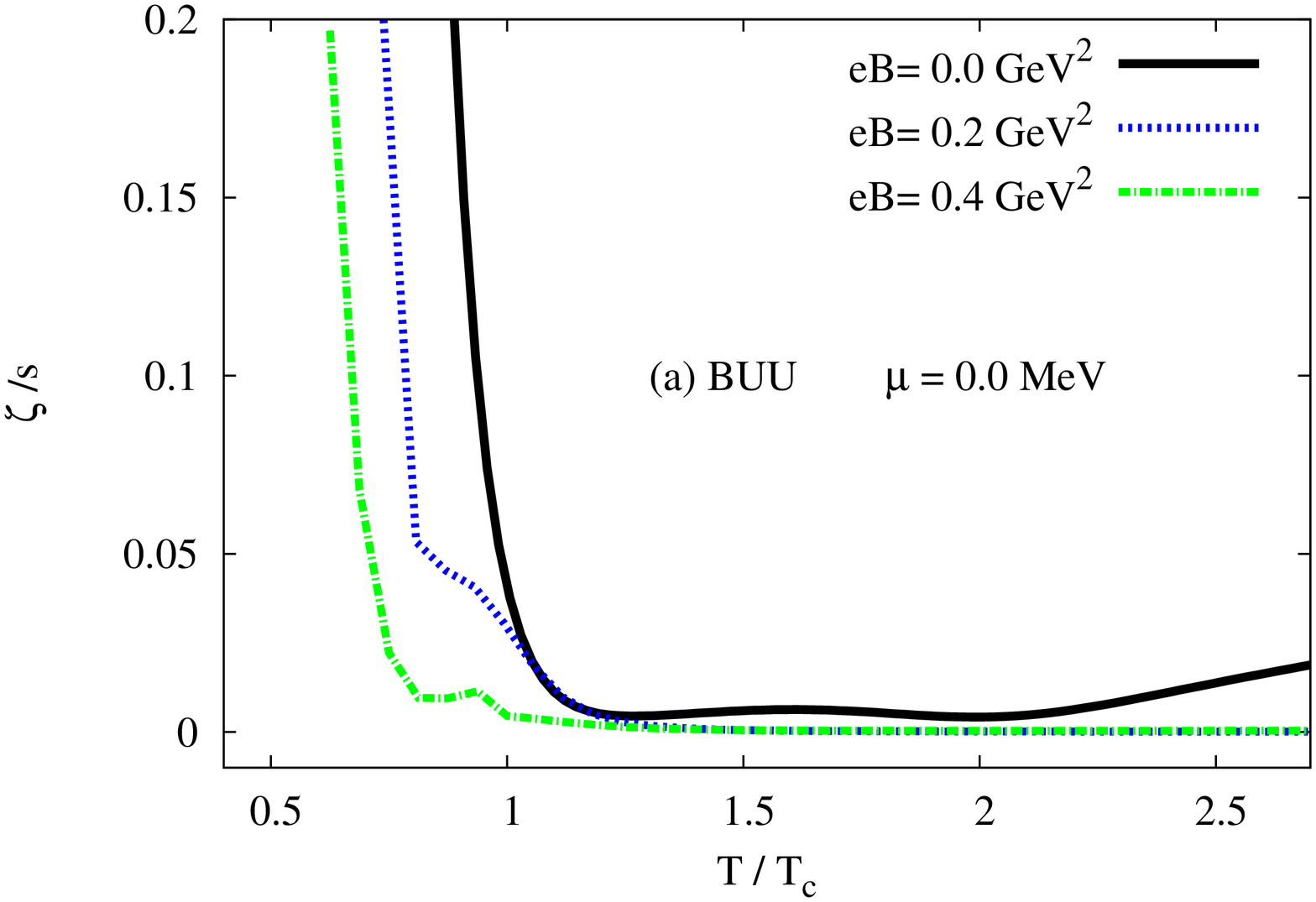}
\includegraphics[width=5.5cm,angle=-90]{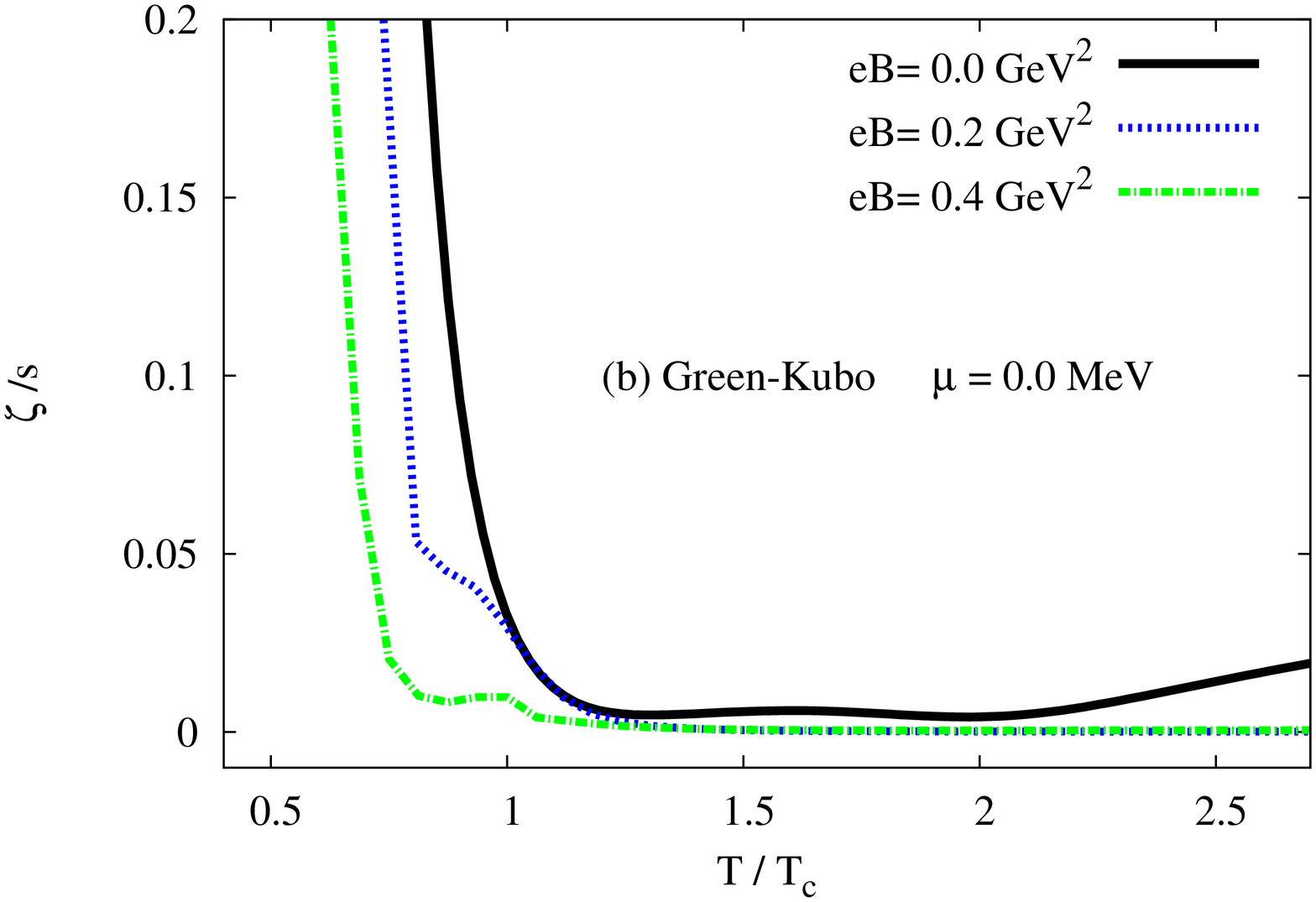}
\caption{\footnotesize (Color online) $\zeta/s$ is illustrated as a function of temperature at vanishing chemical potential and various magnetic field strengths, $eB=0.0$ (solid), $eB=0.2$ (dotted) and $eB=0.4~$GeV$^2$ (dot-dashed curve). Left-hand panel (a) shows results form BUU equation.  The right-hand panel (b) gives the results from GK correlation. \label{fig:Bulk_viscosity2}
}}
\end{figure}

Fig. \ref{fig:Bulk_viscosity2} depicts the influence of finite magnetic field on the temperature dependence of the bulk viscosity normalized to the thermal entropy ($\zeta/s$) at vanishing chemical potential. The solid curve illustrates the results in vanishing magnetic field. The calculations at $eB=0.2$ and $eB=0.4~$GeV$^2$ are presented as dotted  and dot-dashed curves, respectively. The left-hand panel (a) shows $\eta/s$ from Boltzmann-Uehling-Uhlenbeck equation. The right-hand panel is devoted to the same calculations but from Green-Kubo correlation. 

It is obvious that both approaches lead to remarkably almost-identical $\zeta/s$-temperature-dependence. In this regard, even the magnetic field strength does not matter. In both approaches, increasing $e B$ reduces the value of $\zeta/s$, especially at low temperatures. At temperatures exceeding the critical one, the influence of the magnetic field strength drastically reduces. That both BUU and GK produce almost identical  $\zeta/s$ can be understood when comparing Eq. (\ref{eq:buushear}) and Eq. (\ref{eq:gkshear}). Furthermore, GK is based on correlation of the transport coefficients in and out of equilibrium, while BUU is  a generic formalism for all possible interaction in the relativistic system.

It is assumed that, the bulk viscosity can be understood as a conformal equation of state and is a suitable approximation for the weak interaction between quarks and gluons \cite{Arnold:2006}. Furthermore, $\zeta/s$ is believed to draws a picture about massive-to-massless particle ratios. At temperatures exceeding the critical one, we noticed that, $\zeta/s$ infinitesimally decreases with the temperature, especially in nonzero magnetic field. This dependence characterizes a tiny weak coupling between quarks and gluons, where the deconfinement matter becomes dominant. Such negligible monotonic decrease refers to completion of the phase transition from hadrons to quarks.  

Furthermore, we notice that the  magnetic field seems to enhance an appearance of characterizing peaks at the critical temperatures. The peaks are accompanied with minima at low temperatures.

\section{Conclusions \label{conclusion}}

In this paper, we have utilized PLSM with mean field approximation in presence of finite magnetic field in order to address the chiral and deconfinement phase-transitions. We briefly described the structure of  PLSM and shown possible modifications due to finite magnetic field. 

Studying the magnetic field effects on the transport properties such as bulk  ($\zeta$) and shear viscosity ($\eta$), elaborates essential characteristics of the strongly interacting QCD matter and its flow. Both bulk and shear viscosities can be derived from two different approaches. We first utilized the Green-Kubo approach for two-point correlation functions from linear response theory in order to estimate the lowest order of the viscous properties in finite magnetic field. Secondly, we used Boltzmann master equation with Chapman-Enskog expansion in order to derive the relaxation time approximation formulas for both bulk and shear viscosities in nonzero magnetic fields. We conclude that both approaches are almost identical, especially in the hadron phase. This is not the case in the QGP phase. Furthermore, we notice that both quantities (bulk and shear viscosity) are strongly related to the phase transition and how it responses to the instability in the hydrodynamical flow of QGP. Even soft-statistical hadronization leaves fingerprints on bulk and shear viscosity. It is noteworthy mentioning that, they are related to some experimental observables at RHIC and LHC \cite{Torrieri:2009}. At finite  magnetic fields, we have calculated $\zeta/s$ as a function of temperature at vanishing baryon chemical potential. In this regard, we highlight that the speed of sound (or the equation of state) plays an important role in estimating $\eta/s$.

This result confirms the rapid decrease in the chiral phase-transition as well as considerable drop in the critical temperature take place with increasing magnetic field. As the magnetic field increases, a peak appears at the critical temperature.  This can be understood from the  small decrease in the thermal entropy at $T_c$. The latter can be interpreted due to instability in the hydrodynamic flow of QGP and soft statistical hadronization. Also, increasing magnetic field accelerates the transition from hadron to QGP phases, i.e., makes it possible to at lower temperatures.

\appendix 

\section{Viscosity from Boltzmann-Uehling-Uhlenbeck (BUU) equation} 
\label{VBUU6}

The coefficients of the spatial components of the difference between in- and out-of-equilibrium energy-momentum tensor with respect to the Lagrangian density define the transport properties of the system of interest \cite{Tawfik:2010mb}. For an equilibrium state having quark flavors $f$, where  every quark possesses the momentum $\vec{p}$, the phase space distribution is given by $f^{eq}_f$, Eq. (\ref{fqaurk}). For Fermi-Dirac distribution, the symmetric energy-momentum tensor reads \cite{Weinberg1972} 
\bea
T^{\mu\nu} &=& -p\, g^{\mu \nu} + \mathcal{H}\, u^\mu\, u^\nu + \Delta  T^{\mu\nu}, \label{Tensor0}
\eea
where $u^{\nu|\mu}$ being four velocity, $p$ is the pressure, and $\mathcal{H}=p + \epsilon$ is the enthalpy density with $\epsilon = -p + Ts+\epsilon^{\mbox{field}}$ is the energy density including the energy density due to the existence of finite magnetic filed $\epsilon^{\mbox{field}}= eB\cdot \mathcal{M}$ \cite{lattice:2014} and $s$ is the entropy density. When adding a dissipative part $\Delta \, T^{\mu\nu}$ to the energy-momentum tensor, then  
\bea
\Delta \, T^{\mu\nu} &=& \eta \Big( D^\mu u^\nu +  D^\nu u^\mu + \frac{2}{3} \Delta^{\mu\nu} \partial_\sigma u^\sigma \Big) - \zeta \Delta^{\mu\nu} \partial_\sigma u^\sigma, \label{disspative}
\eea
and the Landau-Lifshitz condition, $u_\mu \, \Delta  T^{\mu\nu}=0$ \cite{Weinberg1972}, is satisfied. In local rest-frame, the hydrodynamic expansion reads \cite{Weinberg1972}
\bea
\delta T^{ij } &=&\sum_f \int d\Gamma^* \frac{p^i\, p^j}{E_f}  \Big[ -\mathcal{A}_f \,\partial_\sigma u^\sigma - \mathcal{B}_f\, p_f^\nu D_\nu \left( \frac{\mu}{T}\right) + \mathcal{C}_f\, p_f^\mu  p_f^\nu   \Big( D^\mu u^\nu +  D^\nu u^\mu + \frac{2}{3} \Delta^{\mu\nu} \partial_\sigma u^\sigma \Big) \Big] f_f^{eq}, \label{delTij5}
\eea
where $d\Gamma^*$ stands for generic phase-space, the sum runs over independent contributions from quarks or antiquarks, i.e., assuming point interactions and $\mathcal{A}_f,\, \mathcal{B}_f$ and $\mathcal{C}_f$ are functions depending on momentum $p$.

In the framework of PLSM at nonzero magnetic field and taking into consideration the inverse magnetic catalysis  and by implementing Landau quantization \cite{Shovkovy:2013}, a dimension reduction $d$ to $d-2$ becomes possible and the magnetic field is assumed to affect on a point in the $z$ direction, ${\bf B}=B \,\hat{e}_z$.  Accordingly, the phase space distribution should be modified to Eq. (\ref{phaseeB}) 
\bea
\int d\Gamma^{*}  \equiv \int \frac{d^3k}{(2\pi)^3}  \longrightarrow   \frac{|q_f|B}{2\pi} \sum_\nu \int \frac{dk_z}{2\pi} (2-\delta_{0\nu}). \label{phaseeB1}
\eea

Due to symmetry, the integration over $B_f$ in Eq. (\ref{delTij5}) tends to zero and the derivative in local rest-frame vanishes as well, i.e., $\partial_{k} u_0=0$. Thus the summation over $\mu$ and $\nu$ is equivalent to sum over the spatial indices $\rho$ and $\sigma$, i.e., $p_f^i  p_f^j  p_f^\sigma  p_f^\rho=|p_f|^4 (\delta_{ij} \delta_{\sigma \rho}+\delta_{i\sigma} \delta_{j \rho} +\delta_{i\rho} \delta_{j\sigma}$). Also, in local rest-frame, $p_f =p$.  Equating both Eqs. (\ref{disspative}) and Eq. (\ref{Tensor0}) straightforwardly determines the dissipative parts (bulk and shear, respectively) of the energy-momentum tensor. It is advantageous to work in the local rest frame of the fluid. This leads to the bulk and shear viscosity \cite{Chakraborty:2011}
\bea
\zeta  &=& \frac{1}{3} \sum_f   \frac{|q_f|B}{2\pi} \sum_\nu \int \frac{dk_z}{2\pi} (2-\delta_{0\nu}) \frac{|p|^2}{E_f} f_f \mathcal{A}_f \\ 
\eta   &=& \frac{2}{15} \sum_f  \frac{|q_f|B}{2\pi} \sum_\nu \int \frac{dk_z}{2\pi} (2-\delta_{0\nu}) \frac{|p|^4}{E_f} f_f \mathcal{C}_f. 
\eea

For an out-of-equilibrium state, the four velocity $u^{\mu} (x)$ shouldn't  necessarily remain constant in space and time. When assuming a very small departure from local equilibrium,  
\bea
f_f (x,p) &=& f^{eq} \left(u_i\, p^i/T\right) \Big[1+\phi_f (x,p) \Big], 
\eea 
where
\bea
\phi_f &=& \Big[ -\mathcal{A}_f \,\partial_\sigma u^\sigma - \mathcal{B}_f\, p_f^\nu D_\nu \left( \frac{\mu}{T}\right) + \mathcal{C}_f\, p_f^\mu  p_f^\nu   \Big( D^\mu u^\nu +  D^\nu u^\mu + \frac{2}{3} \Delta^{\mu\nu} \partial_\sigma u^\sigma \Big) \Big].
\eea
In order to determine $\mathcal{A}_f$ and $\mathcal{C}_f$, we use Boltzmann master equation \cite{Chakraborty:2011},
\bea
\frac{\partial f_f (x,t,p)}{\partial t} &=& \left(\frac{\partial}{\partial t} \, + \frac{\partial}{\partial x^i} \frac{\partial x^i}{\partial t}+ \frac{\partial}{\partial p^i} \frac{\partial p^i}{\partial t}  \right) f_f (x,t,p) \equiv {\bf C}\,[f_f].
\eea
The right-hand side gives the collision integral. For collisions $\lbrace i\rbrace\leftrightarrow \lbrace j\rbrace$), the equilibrium distribution functions are identical, i.e., $f_{\lbrace i\rbrace}^{eq} = f_{\lbrace j\rbrace}^{eq}$ \cite{Chakraborty:2011} and the collision integral becomes
\bea
{\bf C} &=&  \sum_{\lbrace i\rbrace\lbrace j\rbrace;f} \sum_\nu  \frac{|q_f|B}{2\pi} (2-\delta_{0\nu}) \frac{1}{S} \int  \Big(\frac{dk_z}{2\pi}\Big)_{\lbrace i\rbrace}\Big(\frac{dk_z}{2\pi}\Big)_{\lbrace j\rbrace}  W(\lbrace i\rbrace |\lbrace j\rbrace) F[f_f].
\eea

 The statistical factor $S$ takes into consideration identical particles in initial state. $F[f_f]$ being Bose-Einstein and Fermi-Dirac distribution functions \cite{Chakraborty:2011}. Because of Landau-Lifshitz condition, some constrains can be added to $\phi_f (x,p)$ so that $|\phi_f| \ll 1$ \cite{Chakraborty:2011}. Furthermore, a particular solution conserving Landau-Lifshitz condition was proposed $\mathcal{A}_f = \mathcal{A}_f^{\mbox{par}} - b E_{b,f}$  \cite{Chakraborty:2011}. Then, bulk and shear viscosity reads   
\bea
\zeta  &=& \frac{1}{3} \sum_f   \frac{|q_f|B}{2\pi} \sum_\nu \int \frac{dk_z}{2\pi} (2-\delta_{0\nu})\left[\frac{|\vec{p}|^2}{3} - c_s^2 E_{B,f}^2 \right]  f_f \mathcal{A}_f^{\mbox{par}}, \\ 
\eta   &=& \frac{2}{15} \sum_f   \frac{|q_f|B}{2\pi} \sum_\nu \int \frac{dk_z}{2\pi} (2-\delta_{0\nu}) \frac{|p|^4}{E_{B,f}} f_f \mathcal{C}_f^{\mbox{par}}. 
\eea

In relaxation time approximation, the phase space distributions of quarks and antiquarks can be replaced by their equilibrium ones; $f = f^{eq}+\delta f$, where $\delta f$ is allowed to be arbitrary infinitesimal, while the collision integral can be given as ${\bf C}_f = \delta f/\tau_f$ \cite{Chakraborty:2011}. Also, the particular solutions $\mathcal{A}_f^{\mbox{par}}$ and $\mathcal{C}_f^{\mbox{par}}$ are given as \cite{Chakraborty:2011},
\bea
\mathcal{A}_f^{\mbox{par}} &=& \frac{\tau_f}{3T} \, \left[\frac{|\vec{p}|^2}{3} - c_s^2 E_{B,f}^2 \right],\\ 
\mathcal{C}_f^{\mbox{par}} &=& \frac{\tau_f}{2T E_f}
\eea 
The bulk and shear viscosities can be reexpressed (for the sake of simplicity, we give the expressions in local rest-frame of the fluid), 
\bea
\zeta (T, \mu, eB) &=&  \frac{1}{9\, T}  \sum_f   \frac{|q_f|B}{2\pi} \sum_\nu \int \frac{dk_z}{2\pi} (2-\delta_{0\nu}) \, \frac{\tau_f}{E_{B,f} ^2} \, \Big[\frac{|\vec{p}|^2}{3} - c_s^2 E_{B,f}^2 \Big]^2 \, f_{f} (T, \mu, eB) ,  \label{eq:buubulk01} \\
\eta (T,\mu, eB) &=&  \frac{1}{15\, T}  \sum_f   \frac{|q_f|B}{2\pi} \sum_\nu \int \frac{dk_z}{2\pi} (2-\delta_{0\nu}) \frac{p^4}{E_{B,f}^2}\,  \tau_f  f_f (T, \mu, eB).
\label{eq:buushear01}
\eea
The distribution function $f_{f}$ is very similar to the equilibrium phase-space distribution function, Eq. (\ref{fqaurk}). Thus, we merely have to replace the dispersion relation $E_f$  by the modified one $E_{B,f}$, Eq. (\ref{eq:moddisp}).

\section{Viscosity from Green-Kubo formalism \label{VGK0}}

In order to derive Eqs. (\ref{eq:gkbulk}) and (\ref{eq:gkshear}) from Green-Kubo formalism, both bulk and shear viscosities are given in  Lehmann spectral representation of the two-point correlation functions as the components of the energy-momentum tensor, such as \cite{Kubo:1957}
\bea
\left(
\begin{array}[c]{c}
\zeta  \\ \eta
\end{array}
\right) 
=
\lim_{\omega\rightarrow 0^+} \lim_{|{\bf p}|\rightarrow 0^+} \frac{1}{\omega}
\left(
\begin{array}[c]{c}
\frac{1}{2} A_{\zeta} (\omega, |{\bf p}|) \\ \frac{1}{20} A_{\eta} (\omega, |{\bf p}|) 
\end{array}
\right),
 \label{eq:matrix_field_A}%
\eea
where $A_{\zeta}$  and $ A_{\eta}$ are spectral functions \cite{Kubo:1957}
\bea
A_{\zeta} (\omega, |{\bf p}|)&=& \int d^4 x\; e^{ip\cdot x} \langle\left[\mathcal{P}(x), \mathcal{P}(0)\right] \rangle, \\ 
A_{\eta} (\omega, |{\bf p}|) &=& \int d^4 x\; e^{ip\cdot x} \langle\left[\pi^{ij}(x), \pi^{ij}(0)\right] \rangle,
\eea
with
\bea
\mathcal{P}(x) &=& -\frac{1}{3} T^{i}_{i} (x) - c_s^2 T^{00} (x), \\
\pi^{ij}(x)  &=& T^{ij} (x) -\frac{1}{3} \delta^{ij} T^{k}_{k} (x),
\eea
and $\langle\left[ \cdots \right] \rangle$ donates an appropriate thermal average. 

Details about the deriving shear viscosity shall be presented (bulk viscosity is very similar). We prove both Eqs. (\ref{eq:gkbulk}) and (\ref{eq:gkshear}). The Matsubara propagators are used in calculating the shear viscosity. The energy-momentum tensor can be expressed in terms of the Lagrangian density 
\bea
T^{\mu \nu}= -g^{\mu \nu} \mathcal{L} + \frac{\partial \mathcal{L}}{\partial (\partial_\mu \Phi)} \partial^{\nu} \Phi.
\eea
For bosons, the viscous stress tensor is entirely determined by the Lagrangian parts which are momentum dependent 
\bea
\pi_{\mu\nu} &=& \left( \Delta_{\mu\nu}  \Delta^{\rho\sigma} -\frac{1}{3} \Delta_{\mu\rho}  \Delta^{\nu\sigma} \right) T^{\rho\sigma},
\eea
where  $\Delta^{\mu\nu} = g^{\mu\nu}-u^\mu u^\nu$. In linear response theory (LRT) know as diagrammatic approach, the impact of the dissipative forces on the energy-momentum tensor can be estimated. It is assumed that these forces are small compared to - the typical energies of the system of interest - a strongly interacting system \cite{Sabyasachi:2014}. The linear response of the microscopic viscous stress-tensor $\pi^{\mu\nu}$ to the dissipative forces enables us to relate the correlation function with the macroscopic (shear) viscosity parameter \cite{Lang:2012Lura}. By denoting the appropriate thermal average of any two-point function as $\langle \cdots \rangle$ and giving it as $2 \times 2$ matrix \cite{Lang:2012Lura}, then, the  two point correlator of viscous stress-tensor becomes 
\bea
\Pi_{ab}(|{\bf p}|) = i \int d^4x\; e^{ip\cdot x} \langle \tau_c \pi_{\mu \nu} (x) \pi^{\mu\nu} (0) \rangle^{ab},
\eea
where $a$, $b\in [1, 2]$ represents the thermal indices of the matrix for $\langle \cdots \rangle$  and $\tau_c$ is the time ordering with respect to a contour in the complex time plane.

\begin{center}
\begin{figure}[htb]
 \begin{tikzpicture}
 \pgfplotsset{every x tick label/.append style={font=\LARGE, yshift=0.5ex}}
  \draw[-latex] [black,thick,line width=1.5pt,domain=-180:-88] plot ({cos(\x)}, {sin(\x)});
 \draw [black,thick,line width=1.5pt,domain=-92:0] plot ({cos(\x)}, {sin(\x)});
   \draw [black,dashed,line width=1.5pt,domain=0:89] plot ({cos(\x)}, {sin(\x)});
	\draw[latex-] [black,dashed,line width=1.5pt,domain=89:180] plot ({cos(\x)}, {sin(\x)});
	 \draw[-latex] [line width=1.5pt, black ]  (1,0) -- (1.8,0);
    \draw [line width=1.5pt, black ] (1.5,0) -- (2.5,0);
    \draw[latex-] [line width=1.5pt, black ]  (-1.8,0) -- (-2.5,0);
    \draw [line width=1.5pt, black ]  (-1,0) -- (-1.9,0);
    \node  at (0,1.3) {$(M=\pi,\sigma)$};
    \node  at (1,0.8) {$(p)$};
    \node at (2.,0.3) {$Q$};
    \node at (2.,-0.3) {$(k)$};
    \node at (0,-1.3) {$Q\;\;(p-k)$};
    \node at (-2.,0.3) {$Q$};
    \node at (-2.,-0.3) {$(k)$};
\end{tikzpicture}  
\caption{\footnotesize A schematic one-loop diagram.  \label{digrame} }
\end{figure}
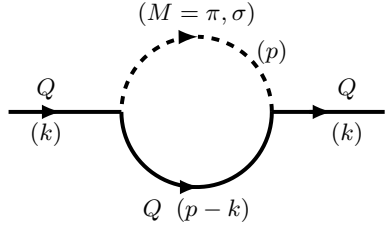
\end{center}

Also, the diagonal element can be related to the retarded two point function of viscous stress-tensor. There are $11$ components corresponding to such functions \cite{Sabyasachi:2014}. The spectral function can be written as
\bea
A_{\eta} (\omega, |{\bf p}|) &=& 2 \tanh \left(\frac{\omega/T}{2}\right)\mbox{Im}\; \Pi_{11} (\omega, p),
\eea
with
\bea
 \Pi_{11}(|{\bf p}|) &=& i \int d\Gamma^{*}  \;N(p,k)\; D_{11}(k)\; D_{11}(p-k), \label{stress}
\eea
and $D_{11}(p)$ is the scalar part of the $11$ components of the quark-propagator matrix and $N(p,k)$ containing the numerator part of two propagators. 
\begin{itemize}
\item The $11$ components of the scalar part of the thermal propagator can be expressed by using the formalism of real-time thermal field theory (RFT) as
\bea
D^{11} (k) &=& \frac{-1}{k_0^2-E_{B,f}^2 + i\epsilon}- 2\pi i\;  E_{B,f}(E_k)\; f_{f}(k)  \delta(k_0^2 - E_{B,f}^2 (k)).
\eea
When replacing the momentum indices $p\rightarrow k$ in Eqs. (\ref{fqaurk}) and  (\ref{fqaurk}), the Fermi-Dirac distribution function and the modified dispersion relation ($f_{f}$ and $E_{B,f}$, respectively) can be reexpressed in finite magnetic field and Polyakov-loop corrections. 

\item What remains in Eq. (\ref{stress}) stands for Fermions \cite{Sabyasachi:2014},
\bea
N(p,k) &=&  \frac{32}{3} k_0 (k_0+\omega) {\bf k} \cdot ({\bf k} +{\bf p}) - 4 \Big( {\bf k} \cdot ({\bf k} +{\bf p}) + \frac{1}{3} {\bf k}^2 ({\bf k} +{\bf p})^2 \Big). \label{Nterm}
\eea
\end{itemize}

Fig. \ref{digrame} illustrates one-loop diagram of quark-meson loops (here $\pi$ and $\sigma$ meson) which can be obtained from the two-point correlation function of the viscous stress-tensor for the quark constituents at the zero frequency and momentum limit \cite{Sabyasachi:2014}. The dashed lines indicates that the quark propagators have an finite thermal width which can be derived from the quark self-energy diagrams.

As an example, we estimate the shear viscosity, Eq. (\ref{eq:matrix_field_A}). The bulk viscosity can be evaluated in a similar manner.  In PLSM in nonzero magnetic field  and by assuming that, the magnetic field is directed along $z$-axis ${\bf B}=B \hat{e}_z$, Eq. (\ref{phaseeB1}), the phase space should be modified according to the magnetic catalysis, Eq. (\ref{phaseeB}). Therefore, The shear viscosity reads 
\bea
\eta &=& \lim_{\omega\rightarrow 0^+} \lim_{|{\bf p}|\rightarrow 0^+} \frac{\mbox{Im} \Pi_{11} (\omega, p)}{10\omega} \\ \nn
&=& \frac{1}{10} \lim_{\omega\rightarrow 0^+} \lim_{|{\bf p}|\rightarrow 0^+} \mbox{Im} \Big[\sum_f \frac{|q_f|B}{2\pi} \sum_\nu \int \frac{dk}{2\pi} (2-\delta_{0\nu}) \frac{(-N)}{E_{B,f}(k) E_{B,f}(p+k)} \lim_{\Gamma\rightarrow 0} \\ \nn && \left( \frac{C^-/\omega}{\left[\omega -E_{B,f}(k) +E_{B,f}(p+k)\right]+i \Gamma}+\frac{C^+/\omega}{\left[\omega +E_{B,f}(k) -E_{B,f}(p+k)\right]+i \Gamma}\right)\Big], 
\eea
where  $C^\mp = \mp{f_{f}(k)}^\mp  + {f_{f}(p+k)}^\mp \left[\mp\omega + E_{B,f}(k)\right]$. $\Gamma$ is the thermal width (or collision rate) of the constituent particles. $\Gamma$ y measures the dissipative coefficients such as the shear viscosity. 

Similar to Ref. \cite{Sabyasachi:2014}, we generalize Eqs. (\ref{eq:gkbulk}) and (\ref{eq:gkshear}), 
\bea
\lim_{p\rightarrow 0} E_{B,f}(p+k)= E_{B,f}(k),
\eea
As given in Ref. \cite{Lang:2012Lura} and by expanding $\Gamma$ in a Laurent series \cite{Lang:2012Lura}, the contribution to the shear viscosity can be given as
\bea
\eta &=& \frac{1}{10}  \frac{|q_f|B}{2\pi} \sum_\nu \int \frac{dk}{2\pi} (2-\delta_{0\nu}) \frac{(-N_0)}{4\,E^2 _{B,f}(k)\, \Gamma}\; \Big[ I^- + I^+ \Big], \; \; \; \mbox{with} \; \; \;  I^\mp = \lim_{\omega\rightarrow0}  \frac{C^\mp (\omega)}{\omega}. 
\eea
Here $I^\mp$ stands for an undefined quantity as $0/0$. Then, we can apply the l'Hospital's rule \cite{Lang:2012Lura}, 
\bea
I^\mp &=& \lim_{\omega\rightarrow0} \frac{\frac{d}{d\omega}\, \left\lbrace C^\mp (\omega)\right\rbrace}{\frac{d}{d\omega} \left\lbrace \omega \right\rbrace} = \frac{1}{T} {f_{f}(k)}^\mp \Big[ 1 +{f_{f}(k)}^\mp \Big].
\eea

The shear viscosity becomes
\bea
\eta &=& \frac{1}{10\, T}  \frac{|q_f|B}{2\pi} \sum_\nu \int \frac{dk}{2\pi} (2-\delta_{0\nu}) \frac{(-N_0)}{4\,E^2 _{B,f}(k)\, \Gamma} {f_{f}(k)}^\mp \Big[ 1 +{f_{f}(k)}^\mp \Big], \label{sheavs4}
\eea
where $N_0= \lim_{\omega,|{\bf p}|\rightarrow 0^+} N\Big(k_0 = \pm E_{B,f}(k), {\bf k}, {\bf p}  \Big)
$. Thus, Eq. (\ref{Nterm}) becomes equivalent to $-14 k^4/3$.  By linking the decay width to the relaxation time, the shear viscosity can be defined in Green-Kubo correlation, Eq. (\ref{eq:gkbulk}), and the shear viscosity reads
\bea
\eta &=& \frac{2}{15\, T}  \frac{|q_f|B}{2\pi} \sum_\nu \int \frac{dk}{2\pi} (2-\delta_{0\nu}) \frac{k^4 \tau_f}{4\,E^2 _{B,f}(k)} {f_{f}(k)}^\mp \Big[ 1 +{f_{f}(k)}^\mp \Big]. \label{sheavs4}
\eea


\end{document}